\documentclass[prb,twocolumn,showpacs,amsmath,amssymb,aps,10pt,superscriptaddress]{revtex4-1}

\usepackage{braket} 
\usepackage{graphicx} 
\usepackage{color}

\usepackage{hyperref}
\hypersetup{%
  colorlinks=true,
  allcolors=blue,
  bookmarksnumbered=true,
}

\begin{document}

\title{Real-space study of the optical absorption in alternative phases of silicon} 
\author{Chin Shen Ong}
\affiliation{Department of Physics, University of California at
  Berkeley and Materials Sciences Division, Lawrence Berkeley National
  Laboratory, Berkeley, California 94720, USA}
\author{Sinisa Coh}
\affiliation{Department of Physics, University of California at
  Berkeley and Materials Sciences Division, Lawrence Berkeley National
  Laboratory, Berkeley, California 94720, USA}
\affiliation{Materials Science and Engineering, Mechanical
  Engineering, University of California Riverside, Riverside, CA
  92521, USA}
\author{Marvin L. Cohen}
\author{Steven G. Louie}
\affiliation{Department of Physics, University of California at
  Berkeley and Materials Sciences Division, Lawrence Berkeley National
  Laboratory, Berkeley, California 94720, USA}
\email{sglouie@berkeley.edu}
\date{\today}

\begin{abstract}
We introduce a real-space approach to understand the relationship between optical absorption and crystal structure.  We apply this approach to alternative phases of silicon, with a focus on the Si$_{20}$ crystal phase
as a case study.  We find that about 83\%\ of the changes in the calculated low-energy absorption in Si$_{20}$ as compared to Si in the diamond structure can be attributed to reducing the differences between the on-site energies of the bonding and anti-bonding orbitals as well as increasing the hopping integrals for specific Si-Si bonds.
\end{abstract}

\maketitle

\section{Introduction} 

In order to reduce the cost of solar-cell energy generation, a great
deal of effort has been put into attempts to increase the number of
charge carriers collected by the solar cell relative to the number of
incident photons (quantum efficiency).  Silicon is the most widely
used photovoltaic material. In terms of global annual power
production, a recent market survey shows that crystalline silicon
dominates the photovoltaic industry 
90\%. One of the major reasons for its popularity is that silicon is
non-toxic and abundant. There are also benefits from technologies
developed over the years in the microelectronics industry.

Despite its widespread usage as a photovoltaic material, silicon does
not efficiently absorb most of the light in the solar spectrum. The
solar spectrum that is received at the Earth's surface (under the
so-called air mass of~1.5 or AM~1.5 for short\cite{astm}) ranges from
0.3~eV~to 4.4~eV and is the strongest around 1.2~eV.  Since silicon
has a direct band gap of 3.3~eV, optical absorption due to direct
transitions can only take place at the high-energy end of the solar
spectrum between 3.3 and 4.4~eV.  Phonon-assisted indirect
transitions\cite{phonon_si,phonon_si_expt} lower the onset of optical
absorption to 1.2~eV. Even then, absorption coefficients due to
indirect transitions alone are smaller and require the solar cell to
be thick in order to amplify the phonon contributions. With a thicker
absorber layer, the solar cell has to have high purity to prolong its
carriers lifetime. Together, the increased thickness and need for
material purity add to the cost of production.

Under ambient conditions, the diamond cubic phase (diamond-Si) is the
most stable crystal phase of silicon, and this is also the crystal
phase of silicon most commonly used to make solar cells today.
However, silicon is known to exist in other crystal phases as well.
For instance, with increase in pressure, silicon undergoes phase
transitions from the diamond-Si phase to the $\beta$-Sn
phase,\cite{bsn} \textit{Imma} phase\cite{imma}, simple hexagonal
phase \cite{simhex1,simhex2,sh_sc1, sh_sc2} and \textit{Cmca}
phase\cite{cmca}. Pressure release from the $\beta$-Sn phase does not
recover the diamond-Si phase. Instead, a slow pressure release
produces the metastable R8 phase\cite{r8} which subsequently
transforms into the BC8 phase,\cite{bc8_0,bc8_1,bc8_2,bm_r8bc8} while
a very rapid pressure release leads to two other tetragonal
phases.\cite{rapidreleasephase} Many of these phases are not suitable
to make solar cells. For example, the first four phases mentioned
above only exist under high pressure. The $\beta$-Sn and simple
hexagonal phases are also metallic\cite{sh_sc1, sh_sc2} while the BC8
phase\cite{bm_r8bc8} is semi-metallic. On the other hand, phases like
the R8\cite{maloner8} and body-centered tetragonal\cite{malonebct}
phases are semiconducting, and since they have direct band gaps
smaller than diamond-Si's, they in principle can also absorb light
over a wider energy range\cite{Cohen2011} than diamond-Si.

One approach\cite{Cohen2011} to increasing the absorption range of
silicon is then to find a crystal phase of silicon that has a smaller
direct band gap than that of diamond-Si. With the advent of
first-principles computational techniques, it has become possible to
search\cite{pso,matfly,airss,csa} for crystal phases that have not
been previously discovered. Botti \textit{et al.}~\cite{botti} found
several crystal phases of silicon that have lower energies than the R8
and BC8 phases and have quasiparticle band gaps ranging from 0.8 to
1.5~eV from GW calculations. Wang \textit{et al.} \cite{wangqq}
proposed phases of silicon that have band gaps from 0.39 to 1.25~eV
obtained within density functional theory (DFT) using the hybrid HSE
functional. Focusing on silicon with direct gaps, Lee \textit{et al.}
\cite{lee_dihedral} presented several other silicon phases.

Recently, Xiang \textit{et al.} in Ref.~\onlinecite{cso_si20} found
the structure of Si$_{20}$ (also called Si$_{20}$-T) using the
particle swarm optimization (PSO)\cite{pso} approach. Their calculated
band gap of Si$_{20}$ is 1.55~eV within DFT-HSE, which is close to the
optimal gap (1.3--1.4~eV)\cite{sq_1.34, sq_quiesser} for solar energy
conversion according to the Shockley-Quiesser model.\cite{sqlimit} One
of the structural features of Si$_{20}$, which is not found in
diamond-Si, is that some of the bonds form equilateral triangles. In
Ref.~\onlinecite{cso_si20}, it was suggested that these bonds might be
related to its improved optical absorption. Nevertheless, the
microscopic reason for the increase in the calculated absorption in
Si$_{20}$ remained unknown.  In a related work, Guo \textit{et al.} in
Ref.~\onlinecite{triangle} proposed an alternative ground state of
silicon with a band gap of 0.61~eV from DFT-HSE that also contains
triangular bonds.

The purpose of this work is to understand how the structure of an
alternative silicon phase may lead to an improved calculated
absorption relative to diamond-Si.  While there are many proposed
metastable phases of silicon with improved absorption, we focus here
on Si$_{20}$ as a case study for our approach since Si$_{20}$ has a
desired calculated optical absorption. (We also note that Si$_{20}$
has a somewhat high formation
energy,\cite{lee_dihedral,commentsi20,commentsi20reply} which may make
it harder to access experimentally.)

One of the obstacles in establishing the relationship between the
crystal structure and optical absorption is the fact that the crystal
structures of Si$_{20}$ and diamond-Si are very different.  For
example, one cannot be related to the other by the removal or addition
of a single atom, or by a small structural distortion that will not
drastically disturb the bonding network of the silicon atoms.
Moreover, the primitive unit cell of diamond-Si contains two atoms
whereas that of Si$_{20}$ contains 20 atoms.  Therefore, a
conventional analysis of optical absorption in the reciprocal space is
non-trivial as each k-point in Si$_{20}$ contains 40 valence and 40
conduction $sp^3$-like bands (unlike diamond-Si, which only has four
of each).

To overcome this difficulty, we study the optical absorption in a real
space representation.  Our real space analysis reveals that about
33\%\ of the enhanced optical absorption of Si$_{20}$ can be
attributed to the decreased differences of the on-site energies
between the bonding and anti-bonding orbitals.  Roughly 50\% is due to
the increased hopping integrals between the bonding and anti-bonding
orbitals.  The remaining 17\% is due to a variety of other
contributions.

\section{Method} 
In this section, we will first describe the conventional density
functional theory (DFT) interband-transition approach and the GW plus
Bethe-Salpeter equation (GW-BSE) approach for computing optical
absorption in reciprocal space.  The latter approach includes electron
self-energy and electron-hole (excitonic) effects. Next we briefly
introduce a real-space representation of the electronic structure in
terms of Wannier functions.  Finally, we transform the expression for
the optical absorption from the reciprocal space representation into
the real space representation.

\subsection{Optical absorption}
 
Optical absorption can be expressed through $\epsilon_2 (\omega)$, the
imaginary part of the dielectric function. Within the
independent-particle DFT approach and neglecting the photon momentum,
the diagonal elements of $\epsilon_2 (\omega)$ can be computed using
the random-phase approximation for a specific light polarization,
\begin{align} 
\begin{split}
\label{eq:abs} 
\epsilon_2 (\omega) =
& 8 \pi^2 e^2 \hbar^2 
\sum_{\mathbf{k}} \sum_{n \in \{C\}}\sum_{m \in \{V\}}|\mathbf{e} 
\cdot \braket{n\mathbf{k}| 
\mathbf{r}|m\mathbf{k}}|^2 \\
&\times \delta(\hbar\omega - E_{n\mathbf{k}} 
+E_{m\mathbf{k}}).
\end{split}
\end{align}
Here $\mathbf{k}$ is the wave vector, $\mathbf{e}$ is the polarization
direction, $\mathbf{r}$ is the position operator, $\omega$ is the
frequency of absorbed photon, $E_{n\mathbf{k}}$ and $E_{m\mathbf{k}}$
are the DFT eigenvalues, $\ket{n\mathbf{k}}$ and $\ket{m \mathbf{k}}$
are the DFT Bloch eigenstates and $\{V\}$ and $\{C\}$ are the valence
and conduction bands. The matrix element $\braket{n\mathbf{k}|
  \mathbf{r}|m\mathbf{k}}$ describes a transition of an electron from
state $\ket{m\mathbf{k}}$ into state $\ket{n\mathbf{k}}$ upon the
absorption of a photon.

\begin{figure}[!t] 
 \includegraphics[scale=1.0]{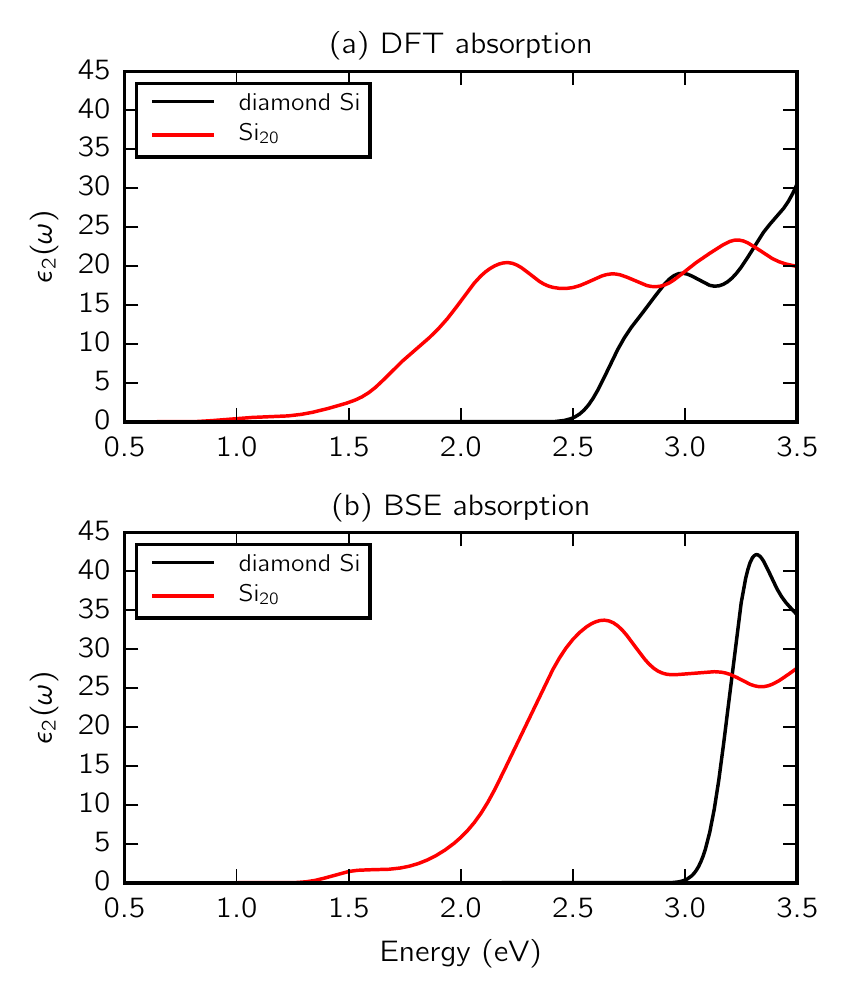}
 \caption{\label{fig:abs} The absorption of diamond-Si (black) and Si$_{20}$ (red)
 calculated with the DFT (a) and BSE (b) approaches. } 
\end{figure}

The $\epsilon_2 (\omega)$ calculated within the DFT approach is shown
in Fig.~\ref{fig:abs}a for diamond-Si (black) and Si$_{20}$ (red).  In
this calculation we used a norm-conserving pseudopotential and we used
the local density approximation as implemented in
Quantum-ESPRESSO.\cite{giannozzi} The plane-wave cutoff for the
electron wavefunction is 36~Ry. For diamond-Si, the Wannier functions
are constructed from a coarse k-mesh of $16\times 16 \times 16$ and
they are used to interpolate quantities on a fine k-mesh of $30 \times
30 \times 30$ to calculate $\epsilon_2 (\omega)$. For Si$_{20}$, the
coarse k-mesh is $8\times 8 \times 8$ and the fine k-mesh is $20
\times 20 \times 20$.
 
From Fig.~\ref{fig:abs}a, it is clear that within the DFT approach,
the onset of optical absorption in Si$_{20}$ is 1.7~eV lower in energy
than in diamond-Si.  However, absorption of Si$_{20}$ at the
absorption edge is relatively small, and it increases significantly
only at 0.8~eV above the absorption edge.  Comparing the steep edges
of the absorption spectra, the steep edge of Si$_{20}$ is still about
0.9~eV lower in energy than it is for diamond-Si.%

In what follows, we discuss two well-known limitations of the optical
absorption calculated within the DFT-RPA approach. The first
limitation is that the calculated DFT-LDA band gap is typically too
small due the fact that DFT eigenvalues are not quasiparticle
excitation energies.  The GW approximation\cite{gw} removes this
limitation by properly including the electron self energy effects.  In
the case of Si$_{20}$ and diamond-Si, the inclusion of the GW
correction\cite{tobepublished} separates the DFT valence and
conduction bands by 0.7--0.8~eV (depending on the k-points and
electron bands) which is close to the value obtained by the
hybrid-functional approach in Ref.~\onlinecite{cso_si20}.

The second limitation of the optical absorption calculated within the
DFT approach is that it does not consider electron-hole
interactions. Within the interacting many-electron picture, an
electron is excited from a ground state $\ket{0}$ to an excited
excitonic state $\ket{S}$ in which the electron interacts with the
hole that it left behind. This process can be
calculated\cite{bsepaper} by solving the BSE and $\epsilon_2 (\omega)$
is then expressed as,
\begin{equation} 
\epsilon_2 (\omega) =8 \pi^2 e^2 \hbar^2  \sum_{S} |\mathbf{e} \cdot\braket{S| \mathbf{r}|0}|^2 
\delta(\hbar\omega - \Omega_S ).
\end{equation} 
Here S labels the exciton states and $\Omega_S$ is the exciton eigenenergy. 

The $\epsilon_2 (\omega)$ spectra calculated\cite{tobepublished} by
solving the BSE for Si$_{20}$ and diamond-Si are shown in
Fig.~\ref{fig:abs}b.  Comparing the GW-BSE and DFT absorption spectra,
we see two main differences.  First, the absorption edge in the GW-BSE
spectrum is 0.6~eV higher in energy than the edge in the DFT spectrum.
This shift is close to the shift resulting from the GW correction
(0.7~eV).  The second difference with the GW-BSE approach is that
$\epsilon_2 (\omega)$ is larger in amplitude by a factor of about
1.5--2.0 near the band edge.

Therefore, while the optical absorption in absolute terms is very
different between the GW-BSE and DFT approaches, the corrections made
by the GW-BSE approach are nearly the same for both Si$_{20}$ and
diamond-Si.  To better understand the improved absorption of
Si$_{20}$, it is sufficient to focus on an analysis of results from
the DFT-RPA approach, since the geometric effect of the crystal
structure is already present at the DFT-RPA level.

\subsection{Localized representation}

The Bloch states appearing in the expression for $\epsilon_2 (\omega)$
(in Eq.~\ref{eq:abs}) have a well-defined crystal momentum
$\mathbf{k}$. They are eigenstates of the Kohn-Sham Hamiltonian,
\begin{equation}
\label{eq:hambloch}
\braket{n\mathbf{k}|H|m\mathbf{k}}=\delta_{nm}E_{n\mathbf{k}}.
\end{equation}
By superposing the Bloch states of different crystal momenta $\mathbf{k}$, 
one can construct a well localized Wannier state, 
\begin{equation} 
\label{eq:b2w} 
\ket{j\mathbf{R}} = \frac{1}{N_\mathbf{k}} \sum_{n\mathbf{k}}
e^{-i\mathbf{k} \cdot \mathbf{R}} U_{nj}^{(\mathbf{k})}\ket{n \mathbf{k}}.
\end{equation} 
Here $\mathbf{R}$ is a real-space lattice vector and
$U_{nj}^{(\mathbf{k})}$ is an arbitrary unitary matrix that mixes the
Bloch bands at $\mathbf{k}$.  In this paper, we use indices $i$ and
$j$ to denote individual Wannier functions and indices $n$ and $m$ to
denote individual Bloch bands.

One often chooses the matrices $U_{nj}^{(\mathbf{k})}$ according to
the scheme introduced by Marzari and Vanderbilt\cite{mv97} so that
$\ket{j\mathbf{R}}$ is as localized in real space around the centers
of mass of the Wannier functions as possible. For this reason,
$\ket{j\mathbf{R}}$ is also called the maximally localized Wannier
function.  The Bloch functions can be reconstructed back from the
Wannier functions through an inverse transformation,
\begin{equation} 
\label{eq:w2b} 
\ket{n \mathbf{k}} = \sum_{j\mathbf{R}} 
e^{i\mathbf{k} \cdot \mathbf{R}} 
U_{nj}^{(\mathbf{k})\dagger} \ket{j\mathbf{R}}. 
\end{equation} 

Since the set of Wannier functions contains the same amount of
information as the set of Bloch bands from which it is generated, it
is convenient to rewrite the Hamiltonian and position operators in the
Wannier basis.  The Hamiltonian in the Wannier (or real space)
representation is simply $\braket{i\mathbf{0}|H|j\mathbf{R}}$ which
can be calculated by a Fourier transform of
$\braket{n\mathbf{k}|H|m\mathbf{k}}$,
\begin{equation}
\label{hamwan}
\begin{split}
\braket{i\mathbf{0}|H|j\mathbf{R}}
&= \frac{1}{N_{\mathbf{k}}} \sum_{nm\mathbf{k}} e^{-i\mathbf{k} \cdot
\mathbf{R}}
U_{ni}^{(\mathbf{k})\dagger}
\braket{n\mathbf{k}|H|m\mathbf{k}}
U_{mj}^{(\mathbf{k})}.
\end{split}
\end{equation}

There are two types of Hamiltonian matrix elements that we will focus
on in this paper.  For the first type, we have $\mathbf{R}=\mathbf{0}$
and $i = j$.  We will refer to this type of matrix element,
\begin{align}
\braket{i\mathbf{0}|H|i\mathbf{0}}=e_{i,}
\label{eq:ei}
\end{align}
 as the 
on-site energy of Wannier function $i$.
The remaining matrix elements 
\begin{align}
\braket{i\mathbf{0}|H|j\mathbf{R}}=t_{ij\mathbf{R}}
\label{eq:tijR}
\end{align}
are known as the hopping integrals.  The hopping integral measures the
probability amplitude for Wannier function $j$ in cell $\mathbf{R}$ to
tunnel to the Wannier function $i$ in the unit cell at the origin.

Wannier functions are constructed from a set of Bloch bands so a
different choice of Bloch bands will lead to different Wannier
functions.  Since the expression for optical absorption in
Eq.~\ref{eq:abs} refers explicitly to occupied and empty Bloch states,
we constructed the Wannier functions either from only empty or only
occupied Bloch states. Therefore, by construction,
$\braket{i\mathbf{0}|H|j\mathbf{R}}$ is zero unless bra and ket are
either both derived from empty or occupied states.

\begin{figure}[!t]
\centering
\sf
(a) Bonding Wannier function

\includegraphics[scale=0.15]{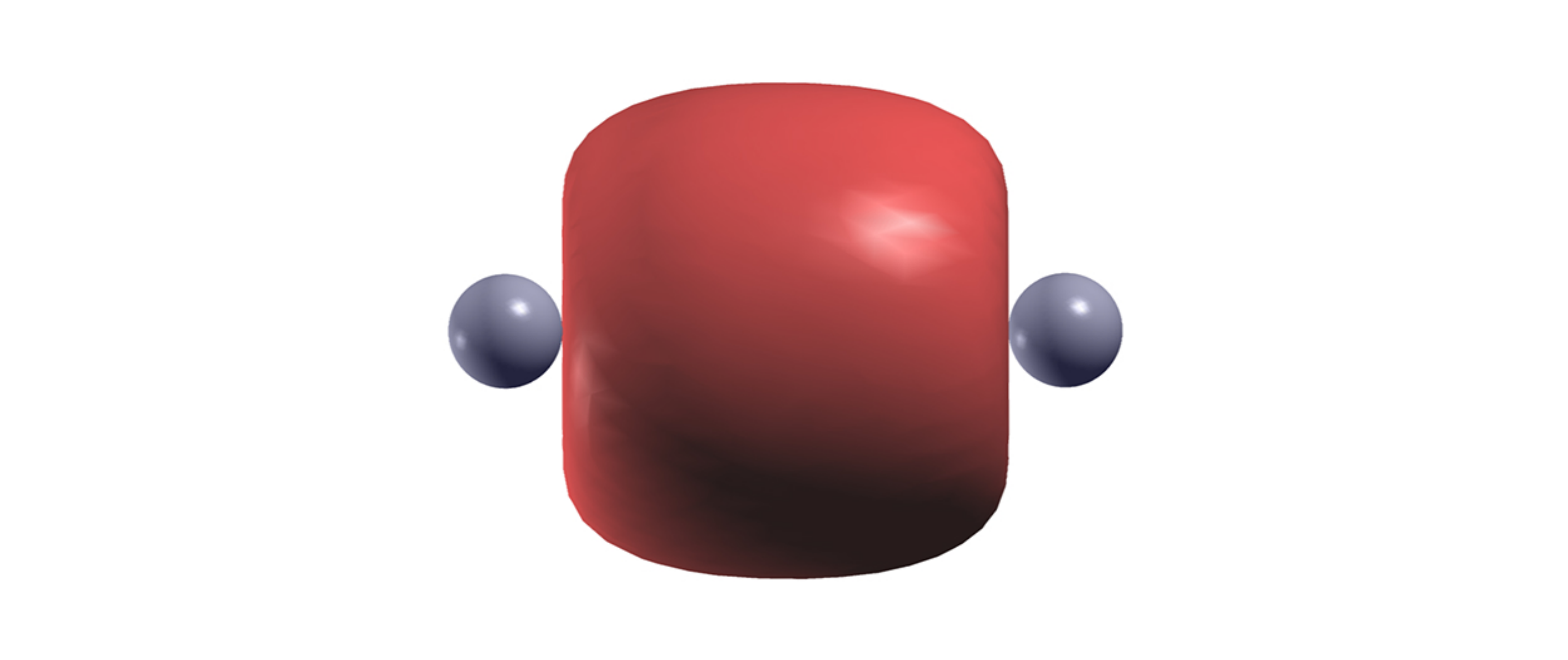}

(b) Anti-bonding Wannier function

\includegraphics[scale=0.15]{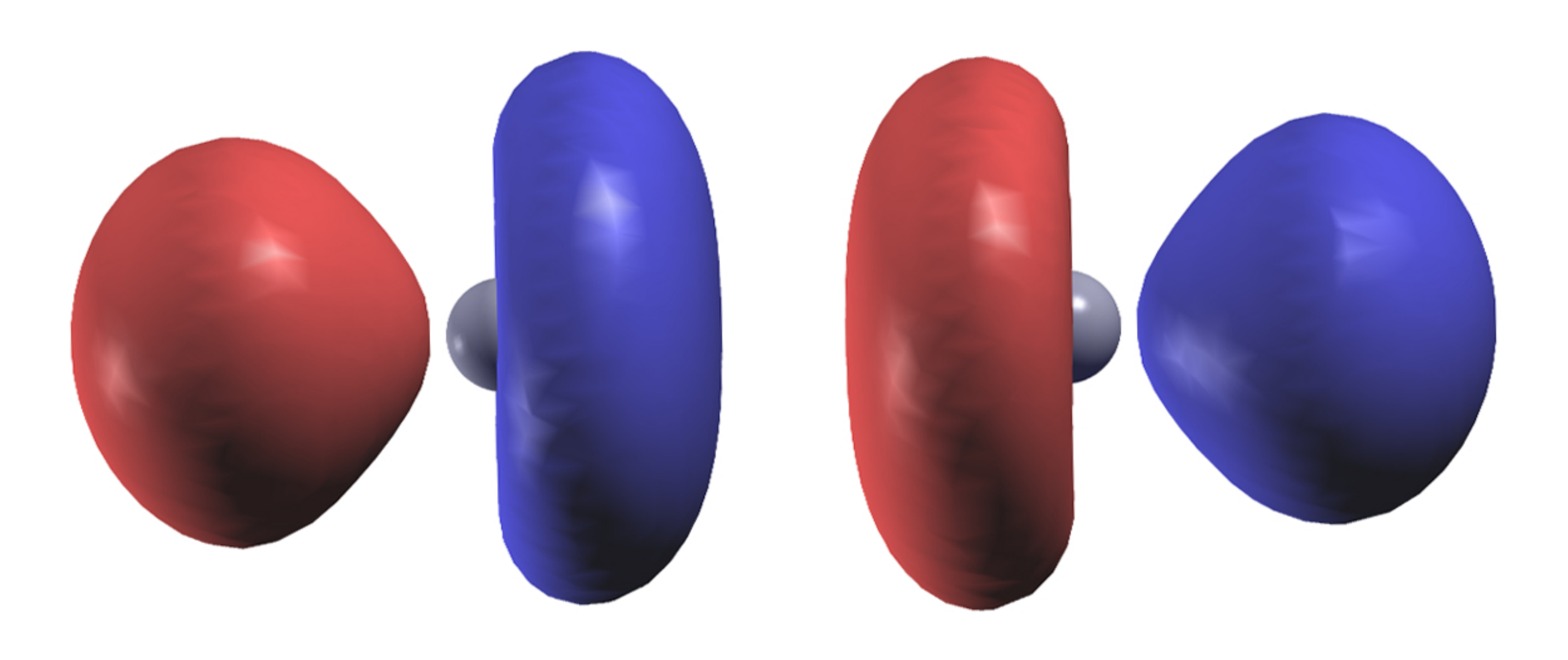}

\caption{\label{fig:wanf}The isosurface of calculated bonding (a) and
  anti-bonding (b) Wannier functions in diamond-Si.  Gray spheres are
  silicon atoms forming the bond.  Isosurface in (a) is 1.4 and 1.0 in
  (b).  Red and blue colors indicate parts of the Wannier function
  with opposite signs.}
\end{figure} 

We will refer to the Wannier functions constructed from the occupied
Bloch states as bonding Wannier functions and those from the empty
states of the relevant conduction bands as anti-bonding Wannier
functions since they typically have real-space forms that resemble
bonding and anti-bonding molecular orbitals. Since silicon bonds are
highly covalent, the valence charges are localized on the bonds
between these two nearest-neighboring silicon atoms. Therefore, the
bonding and anti-bonding Wannier states are localized in the region
between these two silicon atoms, as shown in Fig.~\ref{fig:wanf} for
the case of diamond-Si.  Each Si-Si bond has only one $sp^3$-like
bonding and one $sp^3$-like anti-bonding Wannier function (per each
spin).  For convenience, we will label the on-site energy for the
bonding and anti-bonding states as,
$$ e_i \quad \text{and} \quad \bar{e}_i$$
respectively. Similarly, we denote the hopping integral between anti-bonding states as 
$\bar{t}_{ij\mathbf{R}}$.

\subsection{Optical absorption in the localized basis} 

The optical absorption calculated using $\epsilon_2 (\omega)$
(Eq.~\ref{eq:abs}) within the DFT-RPA approach depends on the energy
of the Bloch states $E_{n\mathbf{k}}$, and the matrix element of the
position operator.  The Bloch state energies are fully determined by
$e_{i}$ and $t_{ij\mathbf{R}}$.  Similarly, the position operator
matrix element can be computed from its representation in the Wannier
basis
\begin{align}
\braket{i\mathbf{0}| \mathbf{r} |j\mathbf{R}} = \mathbf{r}_{ij\mathbf{R}}.
\label{eq:rijR}
\end{align}
In all, optical absorption is exactly determined given the following
three real-space quantities: $e_{i}$, $t_{ij\mathbf{R}}$, and
$\mathbf{r}_{ij\mathbf{R}}$.

\section{Results and discussion}

In this section, we will compare $e_{i}$, $t_{ij\mathbf{R}}$, and
$\mathbf{r}_{ij\mathbf{R}}$ in diamond-Si and Si$_{20}$ and relate
them to the structural differences between the two materials, as well
as the differences in their optical absorption.

\subsection{Comparison of structures}
\label{sec:comp_struct}
 
\begin{figure}[t] 
\includegraphics[scale=0.15]{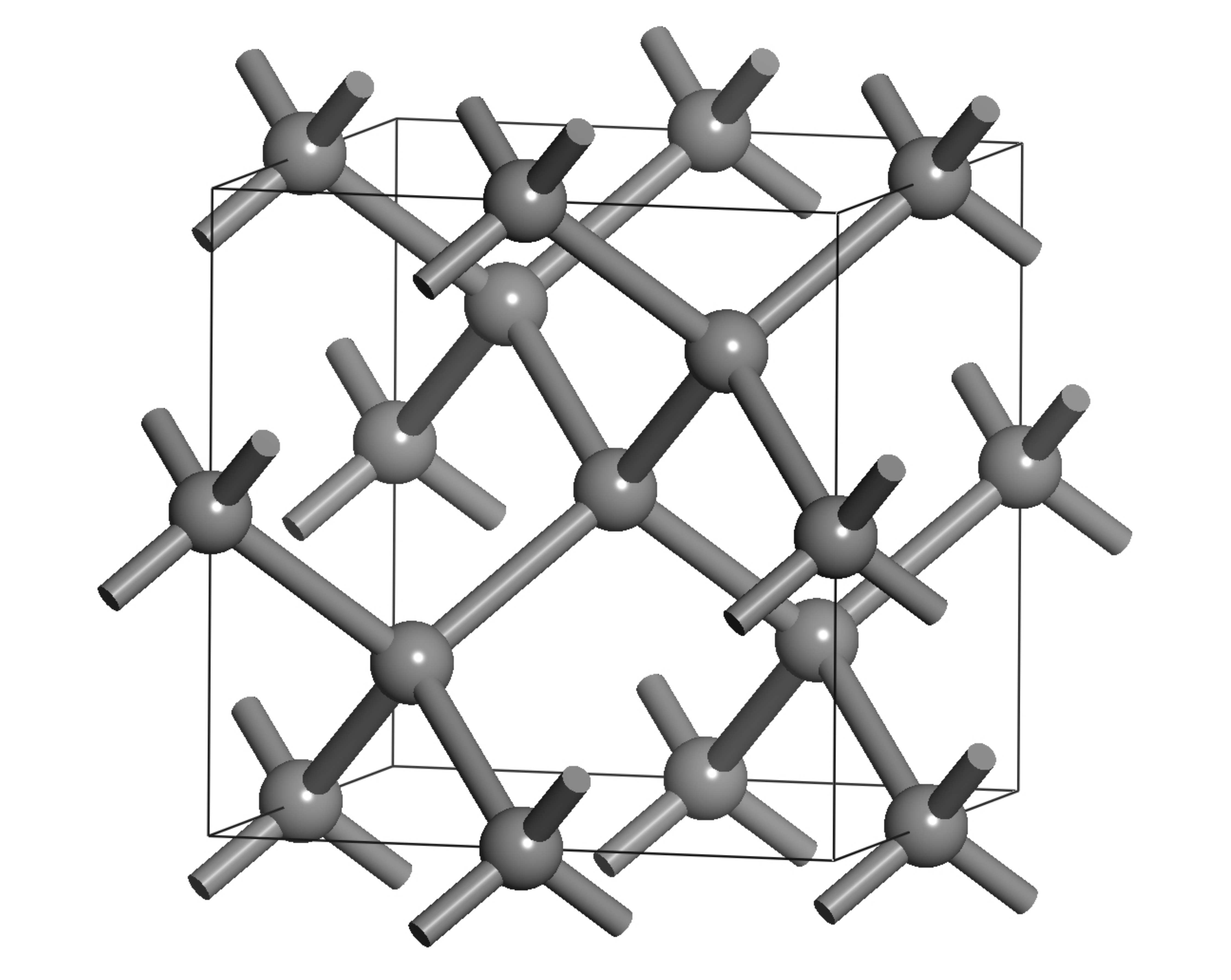}
\caption{\label{fig:structsi2}Conventional unit cell of diamond-Si containing
eight silicon atoms.
 Its primitive unit cell contains only two silicon atoms.} 
\end{figure} 
\begin{figure}[t]
\includegraphics[scale=0.3]{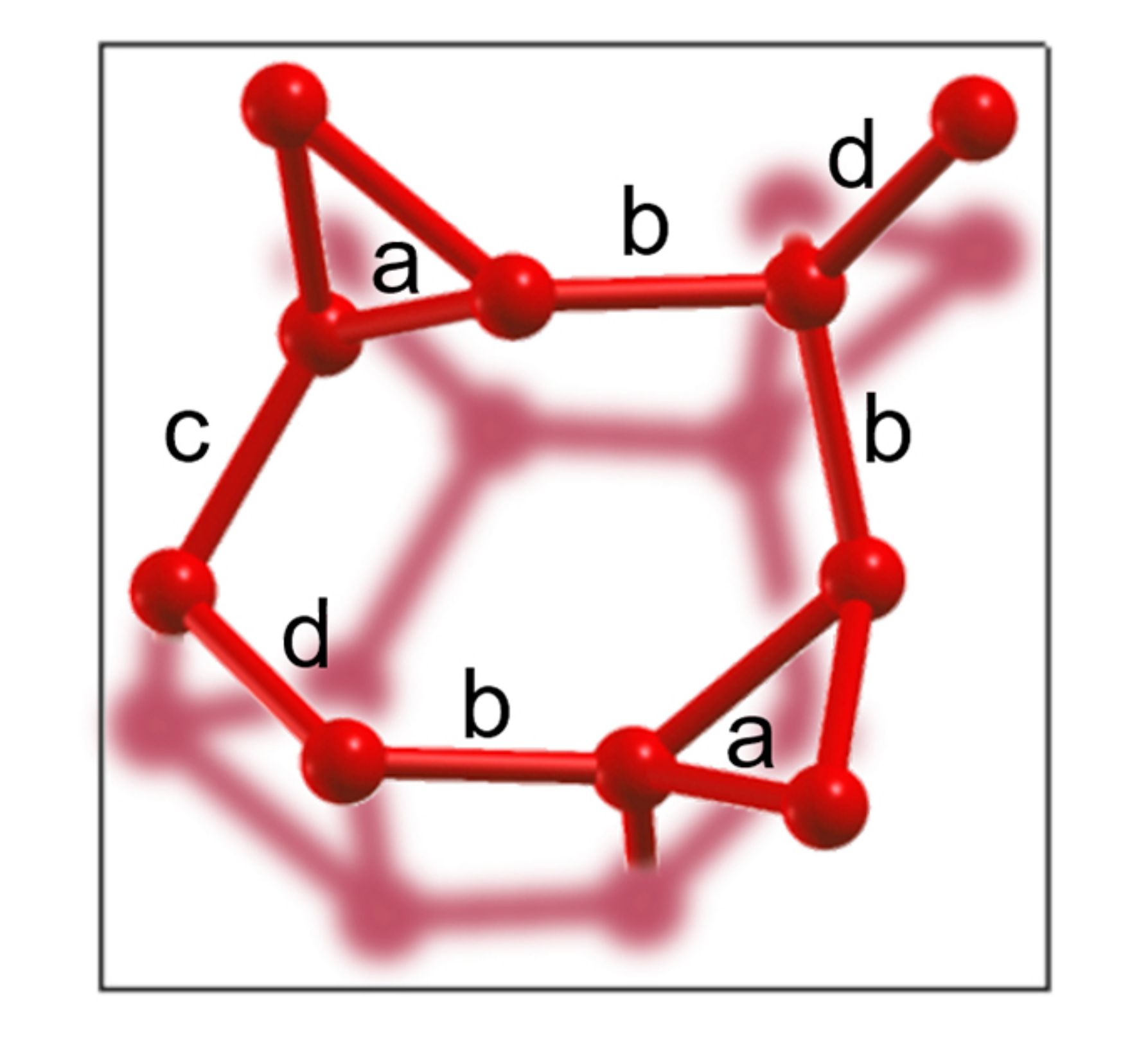}
\caption{\label{fig:structsi20}Conventional unit cell of Si$_{20}$
  containing 20 silicon atoms.  Its primitive unit cell is the same as
  the conventional unit cell. Four distinct Si-Si bonds are indicated
  with labels a, b, c, and d.  Bonds forming a triangle are labelled
  with letter a.}
\end{figure} 

Figures~\ref{fig:structsi2} and \ref{fig:structsi20} show the crystal
structures of diamond-Si and Si$_{20}$.  Both of their conventional
unit cells have cubic lattices. In our calculations, we use fully
relaxed structures of Si$_{20}$ and diamond-Si. The lattice parameters
of the conventional unit cells of Si$_{20}$ and diamond-Si are
7.40~\AA\ and 5.43~\AA. On the average, Si$_{20}$ has one Si atom
every 20.2~\AA$^3$ (2.30~g/cm$^3$) and diamond-Si has one atom every
20.0~\AA$^3$ (2.33~g/cm$^3$).
 
Each Si atom in diamond-Si is tetrahedrally coordinated to four other
Si atoms, such that every bond angle is exactly 109.5$^{\circ}$. Every
Si-Si bond in diamond-Si is symmetrically equivalent. The distance
between the bond centers of two nearest-neighboring bonds is 1.9~\AA.

For Si$_{20}$, every Si atom is also coordinated to four other Si, but
in a distorted tetrahedron. The distortions bring some of the bond
centers of Si$_{20}$ closer together and others further apart.  There
are four symmetry-inequivalent groups of Si-Si bonds in Si$_{20}$ and
they are labelled from a to d in Fig.~\ref{fig:structsi20}.  One
feature of the Si$_{20}$ structure is the type-a bonds which form
triangles. These bonds are highly strained as they are distorted from
109.5$^{\circ}$ to a narrow 60.0$^{\circ}$. As a result, the distance
between two nearest-neighboring bond centers ranges from as short as
1.2~\AA\ (between two type-a bonds of the same triangle) to
2.1~\AA. We will label this range, 1.2--2.1~\AA,\ as the
nearest-neighbor hopping regime.

\subsection{On-site energy $e_{i}$} 
\label{sec:onsite}

Here we compare on-site energies of diamond-Si and Si$_{20}$.  Since 
we can assign a single bonding and anti-bonding Wannier function to
each Si-Si bond, we will focus here on comparing the on-site energies,
$e_i$ and $\bar{e}_i$, for the same bond in the crystal.

\begin{figure}[t]
\includegraphics[scale=1.0]{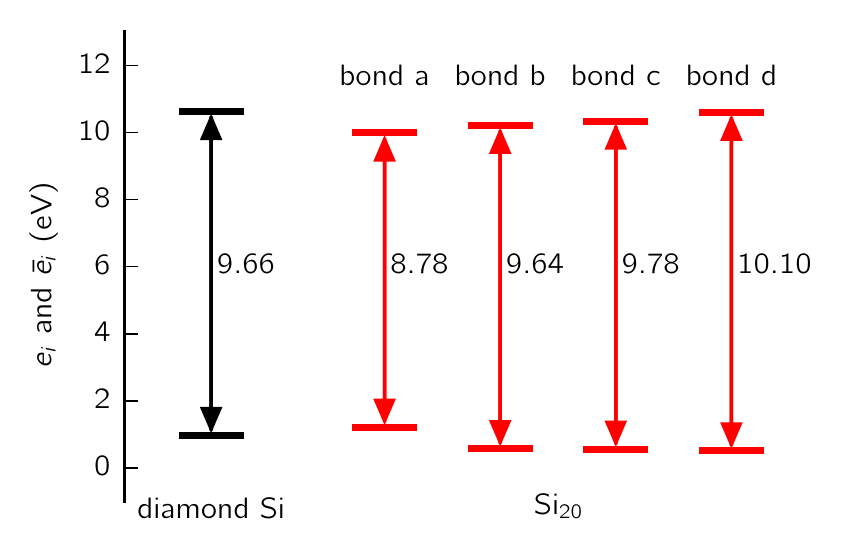}
\caption{\label{fig:onsite}On-site energy of bonding ($e_i$, lower
  value) and anti-bonding ($\bar{e}_i$, higher value) Wannier function
  in diamond-Si (left, black) and Si$_{20}$ (right, red).  Numbers
  indicate $\bar{e}_i-e_i$ in eV.  In the case of Si$_{20}$ we show
  $\bar{e}_i-e_i$ for all four types of bonds.  (The origin of the
  energy scale is arbitrary.)}
\end{figure}

Calculated values of $e_i$ and $\bar{e}_i$ for diamond-Si and
Si$_{20}$ are shown in Fig.~\ref{fig:onsite} with horizontal lines.
The arrow represents the difference between $e_i$ and $\bar{e}_i$ for
a given set of symmetry-related bonds in the structure.  In the case
of diamond-Si, $\bar{e}_i - e_i$ for its Si-Si bond is 9.66~eV.  On
the other hand, $\bar{e}_i - e_i$ for Si$_{20}$ ranges from 8.78 to
10.10~eV.  The smallest value (8.78~eV) belongs to the highly strained
type-a bonds.  Its large deviation from diamond-Si's 9.66~eV is likely
because of its strain, due to the distortion from 109.5$^{\circ}$ to
60$^{\circ}$.  Less strained type-b and type-c bonds have $\bar{e}_i -
e_i$ similar to that in diamond-Si (9.64 and 9.78~eV).  Finally,
type-d bonds have the largest $\bar{e}_i - e_i$ ($10.10$~eV).

We expect that the smaller $\bar{e}_i - e_i$ of type-a bonds would
lower the optical absorption edge of Si$_{20}$ with respect to
diamond-Si's.  This will be analyzed in more detail in
Sec.~\ref{sec:opt_abs}.

\subsection{Hopping integral $t_{ij\mathbf{R}}$}
\label{sec:hopping}

After analyzing $e_i$, we now focus on the hopping integral $t_{ij\mathbf{R}}$
of diamond-Si and Si$_{20}$. 

For the analysis of $t_{ij\mathbf{R}}$, we will define the hopping
distance as the distance between the centers of mass of the Wannier
functions $\ket{i\mathbf{0}}$ and $\ket{j\mathbf{R}}$,
$$\big|\braket{i\mathbf{0}|\mathbf{r}|i\mathbf{0}}-\braket{j\mathbf{R}|\mathbf{r}|j\mathbf{R}}\big|.$$
In what follows, we will relate $t_{ij\mathbf{R}}$ with its hopping distance.

\subsubsection{Bonding states}
\label{hop_b}

\begin{figure}[t]
\includegraphics[scale=1.0]{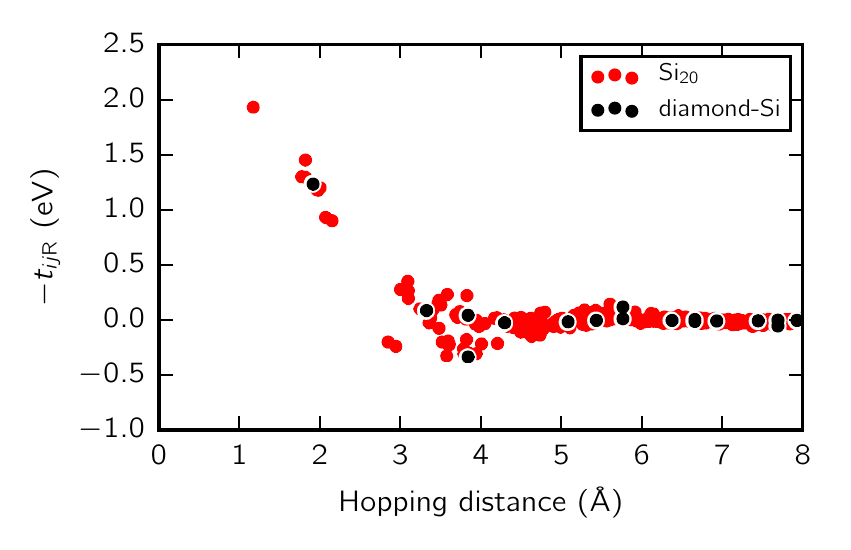}
\caption{\label{fig:hopval} Hopping integrals between bonding Wannier
  functions, as a function of hopping distance for diamond-Si (black)
  and Si$_{20}$ (red).}
\end{figure}

First, we discuss the hopping integrals between bonding Wannier
functions. As shown in Fig.~\ref{fig:hopval}, the hopping integrals of
both diamond-Si and Si$_{20}$ are nearly zero for hopping distances
beyond 5~\AA.  This behavior is characteristic of the exponential
localization\cite{wan_expdecay_2} of Wannier functions for insulators.

The hopping integral $t_{ij\mathbf{R}}$ with the largest magnitude for
diamond-Si is $-1.23$~eV. This hopping integral couples a bonding
Wannier function with its nearest bonding neighbor and has a hopping
distance of 1.9~\AA.  In Fig. \ref{fig:hopval}, it is denoted by the
leftmost black dot.  For Si$_{20}$, hopping integrals coupling the
nearest bonding neighbors are distributed over the range of 1.2--2.1
\AA\ (see Sec.~\ref{sec:comp_struct}).  In Fig.~\ref{fig:hopval}, they
are represented by the group of red dots surrounding the
above-mentioned black dot.

The largest $|t_{ij\mathbf{R}}|$ for Si$_{20}$ corresponds to the
hopping integral with the shortest hopping distance of 1.2~\AA. This
hopping integral couples type-a bonds and is 0.70~eV larger than the
largest $|t_{ij\mathbf{R}}|$ of diamond-Si.  The presence of this
large hopping integral in Si$_{20}$ is due to the fact that the
distance between triangular bonds is $1.9-1.2=0.7$~\AA\ shorter than
the shortest bond--bond distance in diamond-Si.

As we will analyze later in more detail, we expect the larger hopping
integrals of the occupied Wannier functions to raise the valence band
edge in Si$_{20}$ as we expect the valence bands to have a larger
bandwidth.

\subsubsection{Anti-bonding states}
\label{hop_ab}

\begin{figure}[t]
\includegraphics[scale=1.0]{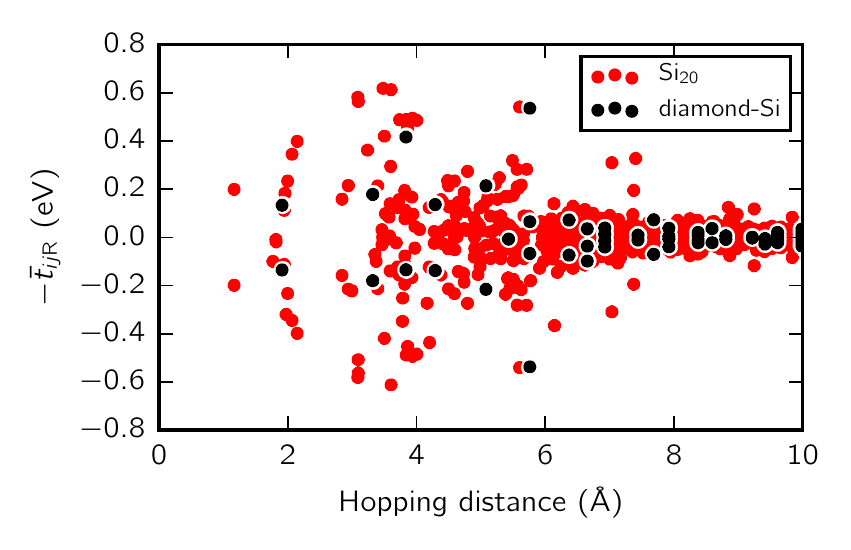}
\caption{\label{fig:hopcond}Hopping integrals between anti-bonding
  Wannier functions, as a function of hopping distance for diamond-Si
  (black) and Si$_{20}$ (red).}
\end{figure}

Now, we look at the hopping integrals between the anti-bonding states.
Figure~\ref{fig:hopcond} shows that the largest
$|\bar{t}_{ij\mathbf{R}}|$ for diamond-Si is 0.54~eV and has a hopping
distance of 5.8~\AA. Unlike the bonding states, this largest
$\bar{t}_{ij\mathbf{R}}$ does not couple the nearest-neighboring
Wannier functions. That hopping integral is four times smaller
(0.13~eV).  For Si$_{20}$, the largest $|\bar{t}_{ij\mathbf{R}}|$ is
0.62~eV and has a hopping distance of 3.5~\AA. It is somewhat larger
than diamond-Si's largest $|\bar{t}_{ij\mathbf{R}}|$ and it also does
not couple the nearest-neighboring Wannier functions.

Nevertheless, in the nearest-neighbor hopping regime of 1.2--2.1~\AA,
the largest $|\bar{t}_{ij\mathbf{R}}|$ in Si$_{20}$ is 0.40~eV. This
value is significantly larger than the corresponding
$|\bar{t}_{ij\mathbf{R}}|$ for diamond-Si (0.13~eV) in the same
regime.

Notably, even though $|\bar{t}_{ij\mathbf{R}}|$ for anti-bonding
Wannier functions are nearly zero above hopping distance of 9~\AA, it
does not increase monotonically below 9~\AA\ as the hopping distance
decreases. The distribution of $\bar{t}_{ij\mathbf{R}}$
(Fig.~\ref{fig:hopcond}) is more dispersive than that of
$t_{ij\mathbf{R}}$ (Fig.~\ref{fig:hopval}). This is likely related to
the fact that the anti-bonding Wannier functions
(Fig.~\ref{fig:wanf}b) have more nodes than the bonding Wannier
functions (Fig.~\ref{fig:wanf}a). They are also more diffuse than the
bonding Wannier functions. In addition, anti-bonding Wannier functions
hybridize with the continuum, making them somewhat sensitive to the
choice of the frozen window used in the Wannier
disentanglement\cite{disentanglement} procedure. (For consistency, we
have chosen the frozen windows in both diamond-Si and Si$_{20}$ to
span from the conduction band minimum (CBM) to 3.7~eV above the CBM.)

Hopping integrals between anti-bonding Wannier states of Si$_{20}$ are
distributed over a wider energy range than diamond-Si.  We expect the
larger hopping integrals between the empty Wannier functions of
Si$_{20}$ to increase the bandwidth of the conduction bands and lower
its lower band edge. This will be further discussed in
Sec.~\ref{sec:opt_abs}.
 
\subsection{Position integral $\mathbf{r}_{ij\mathbf{R}}$} 
\label{sec:position}

Now we discuss the third real-space object required to compute the
optical absorption: position operator in the real space
representation, $\mathbf{r}_{ij\mathbf{R}}$, between a bonding Wannier
function and an anti-bonding Wannier function.  (The matrix elements
between two bonding or two anti-bonding Wannier functions do not enter
into Eq.~\eqref{eq:abs}.)

For diamond-Si, $|\mathbf{r}_{ij\mathbf{R}}|^2$ is the largest when
$i$ and $j$ are both centered on the same bond, as can be
expected. Its value is $|\mathbf{r}_{ij\mathbf{R}}|^2=0.59$~{\AA}$^2$
and it is seven times as large than that between the neighboring bonds
(0.09~{\AA}$^2$).  For Si$_{20}$, the largest
$|\mathbf{r}_{ij\mathbf{R}}|^2$ are also on the same bond.  Their
values for four types of Si$_{20}$ bonds are nearly the same.  Their
average value is 0.53$\pm$0.02~{\AA}$^2$. (The next largest value is
only 0.15~{\AA}$^2$.)

Here, two observations can be made. First, we see that in the real
space representation, $|\mathbf{r}_{ij\mathbf{R}}|^2$, like the
Hamiltonian, is highly localized.  Second, the largest
$|\mathbf{r}_{ij\mathbf{R}}|^2$ for Si$_{20}$ and diamond-Si have
nearly the same numerical value. This is likely because the Wannier
functions of Si$_{20}$ have similar real-space character as those in
diamond-Si.

\subsection{Relating $e_i$ and $t_{ij\mathbf{R}}$ to the optical absorption} 
\label{sec:opt_abs}

\begin{figure}[t]
\includegraphics[scale=1.0]{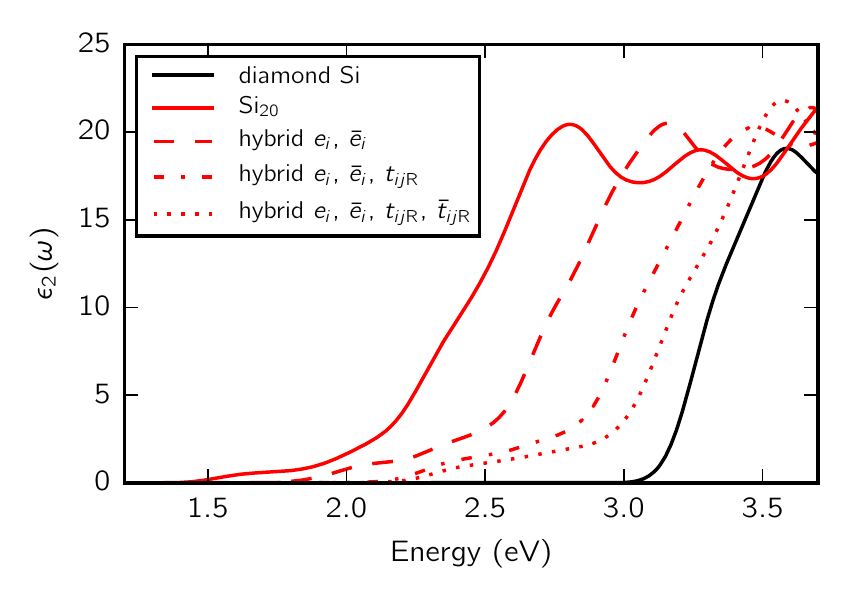}
\caption{\label{fig:abs_mod} Optical absorption in diamond-Si (black),
  Si$_{20}$ (solid red), and two hybrid cases (dashed, dotted, see
  text for details).  Absorption curves are scissor shifted by 0.6~eV
  in all four cases based on our GW-BSE calculation.}
\end{figure}

Now, we will relate the magnitudes of $e_{i}$ and $t_{ij\mathbf{R}}$
to the optical absorption in diamond-Si and Si$_{20}$.  For this
purpose, we compute the optical absorption in three model systems,
which are hybrids between diamond-Si and Si$_{20}$.  These hybrid
systems have the same Hamiltonian as Si$_{20}$, except for some
$e_{i}$, $\bar{e}_i$, $t_{ij\mathbf{R}}$ and $\bar{t}_{ij\mathbf{R}}$
which are modified to resemble those in
diamond-Si. Figure~\ref{fig:abs_mod} shows the calculated optical
spectra of diamond-Si (in solid black), Si$_{20}$ (in solid red), and
the hybrid systems (in dashed, dotted-and-dashed, and dotted red).

The dashed red curve in Fig.~\ref{fig:abs_mod} shows the calculated
optical absorption of the first hybrid system, where all on-site
energies, $e_{i}$ and $\bar{e}_i$, of Si$_{20}$ are made to be equal
to those of diamond-Si.

The dotted-and-dashed curve in Fig.~\ref{fig:abs_mod} represents the
second hybrid system where, on top of the modifications made for the
first hybrid system, hopping integrals $t_{ij\mathbf{R}}$ between
bonding Wannier functions are modified as well. This modification is
done in the following way.  First, we identify hopping integrals in
Si$_{20}$ larger than the nearest-neighbor hopping integral in
diamond-Si (their values are $-1.93$, $-1.45$, $-1.30$, and
$-1.29$~eV).  Second, we modify these hopping integrals so that they
are equal to the nearest-neighbor hopping integral in the diamond-Si
($-1.23$~eV).

Finally, the dotted red curve in Fig.~\ref{fig:abs_mod} shows the
optical absorption of the third hybrid system which, in addition to
the modifications made for the first and second hybrid system, has
modified hopping integrals between the anti-bonding Wannier functions,
$\bar{t}_{ij\mathbf{R}}$.  Here we follow the same logic as is used
for hopping integrals between the bonding Wannier functions.  We first
identify hopping integrals in Si$_{20}$ in the nearest-neighbor regime
that are larger than the nearest-neighbor hopping integral in
diamond-Si (their magnitudes are 0.18, 0.20, 0.23, 0.32, 0.34, and
0.40 eV).  Next, we modify these hopping integrals to the
nearest-neighbor hopping integral between anti-bonding states in
diamond-Si (0.13~eV).

As can be seen from Fig.~\ref{fig:abs_mod}, modifying only $e_i$ and
$\bar{e}_i$ shifts the leading edge of the absorption spectrum of
Si$_{20}$ to a higher energy by about 0.30~eV. This is about 33\%\ of
its difference with diamond-Si.  Modifying $e_i$, $\bar{e}_i$, and
$t_{ij\mathbf{R}}$ further shifts the leading edge of the absorption
spectrum by another 0.30~eV. When $e_i$, $\bar{e}_i$,
$t_{ij\mathbf{R}}$ and $\bar{t}_{ij\mathbf{R}}$ are all modified, the
edge of the absorption spectrum is shifted by a total of 0.75~eV from
the original calculated spectrum which accounts for approximately 83\%
of its difference with diamond-Si.

This behavior can be understood by considering a simple tight-binding
model of a periodic one-dimensional mono-atomic chain.  The band
structure of such a model is given by $e+2t\cos(ka)$ where $e$ is the
on-site energy, $t$ is the hopping integral between the
nearest-neighboring orbitals, and $a$ is the distance between atoms.
Therefore, on-site energy $e$ can be thought of as an average energy
of the band while the hopping integral $t$ determines its bandwidth.
This means that smaller $\bar{e}_{i}-e_{i}$ and larger
$t_{ij\mathbf{R}}$ and $\bar{t}_{ij\mathbf{R}}$ found in Si$_{20}$
will lower the average band gap.

Interestingly, the steep edges of the four absorption curves in
Fig.~\ref{fig:abs_mod} are nearly shifted by the same amount.  This is
consistent with the fact that the position matrix elements do not
change much between the different structures. Instead, the different
spectra mostly result from the different on-site energies and hopping
integrals.
  
The modifications that are made to the hybrid systems do not account
for the remaining 17\% and an absorption tail at low energy.  This can
be attributed to the following two simplifications.  First, we only
modified some of the larger hopping integrals in our calculations of
the hybrid models.  Second, even~though we modified the hopping
integrals in our calculations, we have always kept the structure of
Si$_{20}$ the same.  Therefore, for a given Bloch state, relative
phases between its amplitude and those of its neighboring bonding
sites will still be different from diamond-Si.  In other words, even
if the hopping integrals were somehow made exactly the same in the two
structures, their optical absorption edges may still not be the same
because of this effect.

\section{Conclusion}
The different structure of Si$_{20}$, relative to diamond-Si, leads to
smaller on-site energy differences and larger hopping integrals
between some of its Wannier functions. We have identified that most of
these differences are due to the strained bonds forming triangles
(i.e. type-a bonds) in Si$_{20}$. Different on-site energies and large
hopping integrals are responsible for approximately 83\% of the
improved optical absorption in Si$_{20}$ for photovoltaic applications
relative to diamond-Si. The remaining difference is attributed to
contributions from the smaller hopping integrals and the relative
phase changes in the electron wavefunctions.

Introducing strain to the bonds in the crystal structure turns out to
be important when looking for crystal phases of silicon that have band
gaps smaller than diamond-Si. However, as strain may reduce the band
gap of diamond-Si, it also reduces the stability of the crystal
structure. It is possible that a large band gap reduction may require
a strain that is too large for the crystal structure to be
thermodynamically stable. Hence, in the search for a practically
viable silicon crystal phase that has a band gap smaller than that of
diamond-Si, it is a balance between reducing the band gap and
increasing the strain in the crystal structure.

\begin{acknowledgments}
This work was supported by National Science Foundation Grant
No. DMR-1508412 which provided for the DFT calculations and the
Wannier functions analysis, and by the Theory of Materials Program at
the Lawrence Berkeley National Lab funded by the Director, Office of
Science and Office of Basic Energy Sciences, Materials Sciences and
Engineering Division, U.S. Department of Energy under Contract
No. DE-AC02-05CH11231 which provided for the GW-BSE calculations.
Computational resources have been provided by the DOE at Lawrence
Berkeley National Laboratory's NERSC facility. C.S.O. acknowledges
support from the Singapore National Research Foundation (Clean Energy)
PhD Scholarship.
\end{acknowledgments}

\bibliography{pap}
\bibliographystyle{apsrev4-1}

\end{document}